\documentclass[sigconf]{acmart}

\usepackage{booktabs} 

\setcopyright{rightsretained}

\usepackage{booktabs} 
\newcommand{\overbar}[1]{\mkern 1.5mu\overline{\mkern-1.5mu#1\mkern-1.5mu}\mkern 1.5mu}
\usepackage{graphicx}
\usepackage{listings}
\usepackage{amsmath}
\usepackage{enumitem}
\usepackage{multirow}
\acmConference[PROMISE'17]{The 13th International Conference on Predictive Models and Data Analytics in Software Engineering}{November 2017}{Toronto, Canada} 
\acmYear{2017}
\copyrightyear{2017}

\begin{document}
\title{Scripted GUI Testing of Android Apps:\\A Study on Diffusion, Evolution and Fragility}

\author{Riccardo Coppola, Maurizio Morisio and Marco Torchiano}
\affiliation{%
  \institution{Dipartimento di Informatica e Automatica\\Politecnico di Torino}
  \city{Turin} 
  \state{Italy} 
}
\email{name.surname@polito.it}

\begin{abstract}
\textbf{Background.} Evidence suggests that mobile applications are not thoroughly tested as their desktop counterparts. In particular GUI testing is generally limited. Like web-based applications, mobile apps suffer from GUI test fragility, i.e. GUI test classes failing due to minor modifications in the GUI, without the application functionalities being altered.

\textbf{Aims.} The objective of our study is to examine the diffusion of GUI testing on Android, and the amount of changes required to keep test classes up to date, and in particular the changes due to GUI test fragility. We define metrics to characterize the modifications and evolution of test classes and test methods, and proxies to estimate fragility-induced changes.

\textbf{Method.} To perform our experiments, we selected six widely used open-source tools for scripted GUI testing of mobile applications previously described in the literature. We have mined the repositories on GitHub that used those tools, and computed our set of metrics.

\textbf{Results.} We found that none of the considered GUI testing frameworks achieved a major diffusion among the open-source Android projects available on GitHub. For projects with GUI tests, we found that test suites have to be modified often, specifically 5\%-10\% of developers' modified LOCs belong to tests, and that a relevant portion (60\% on average) of such modifications are induced by fragility.

\textbf{Conclusions.} Fragility of GUI test classes constitute a relevant concern, possibly being an obstacle for developers to adopt automated scripted GUI tests. This first evaluation and measure of fragility of Android scripted GUI testing can constitute a benchmark for developers, and the basis for the definition of a taxonomy of fragility causes, and actionable guidelines to mitigate the issue.
\end{abstract}

%
%


\copyrightyear{2017} 
\acmYear{2017} 
\setcopyright{acmcopyright}
\acmConference{PROMISE }{November 8, 2017}{Toronto, Canada}\acmPrice{15.00}\acmDOI{10.1145/3127005.3127008}
\acmISBN{978-1-4503-5305-2/17/11}

\maketitle

\section{Introduction}

Android has reached a very significant market share with respect to other mobile systems (86.2\% in Q2 '16\footnote{https://www.statista.com/statistics/266136/global-market-share-held-by-smartphone-operating-systems/}), and mobile devices have largely overtaken desktop ones in terms of shipped units (1.91 to 0.25 billion in 2015\footnote{http://www.gartner.com/newsroom/id/3187134}). Mobile devices offer their users a large number of applications, capable of performing tasks that just a few years ago were exclusively available on high-end desktop computers.

One of the characteristics that have brought Android to its success is the availability of marketplaces (e.g., the Play Store) where developers can sell -or release for free- their applications. The huge quantity of software published on those platforms, and the resulting competition, makes crucial for the applications to behave as promised to their users. 

Thus, testing applications and their GUI (i.e., Graphical User Interface), through which most of the interaction with the final user is performed, becomes a valuable practice to ensure that no crashes and no undesired behaviours happen during a typical execution.

However, there is evidence that Android applications -and mobile applications in general- are not deeply tested as they should be. Although a variety of testing tools (open-source or not) are available, most Android developers rely just on manual testing
. Some developers do not perform testing at all, leaving the recognition of faults and bugs to the feedback of their users. Evidence about this lack of testing is given in \cite{understanding-culture}, where only 14\% of the set of applications considered featured any kind of test classes.

In addition to this need for testing, Android development comes with a set of domain-specific challenges, that have consequences for testing. The main differences between traditional software and Android applications are: the great quantity of different context events to which the apps have to react properly; the diversity of devices and configurations where apps will eventually be deployed; the very fast pace of evolution of the operating system; the lack of resources that has been intrinsic for a long time for mobile devices\cite{muccini-1}.

Among those peculiarities of Android testing, we focus on the problem of fragility of GUI test classes. 
We consider the fragility of test classes among the main factors that may discourage the adoption of GUI testing for Android applications, since developers may decide to not exercise any testing if even small changes in the user interface may break entire test suites. 
In our previous study \cite{coppola-1} on a popular open-source Android application, K-9 mail, a small test suite was developed and adapted to different releases. We found that up to 75\% of the tests developed had to be modified because of modifications in the GUI.


With this work we aimed at gathering information about test suites in released open-source projects. We collected statistics about the level of penetration of six popular tools that are used for Android GUI testing, among open-source applications whose source code is available on GitHub. For those projects that leveraged the tools, we measured the quantity of test code they featured, and counted the modifications performed on test classes during their lifespan. 

We define the concept of fragility of test classes, and provide metrics to estimate the fragility of a project by automated inspection of its test suite. This allows us to give a characterization and quantification of the fragility issue on a large set of Android projects, and can be an aid to developers to evaluate the maintenance effort needed by their scripted test suites. This evaluation can serve, in the future, as a base for a taxonomy of fragility causes, a set actionable guidelines to help developers to avoid it, and finally automated tools capable of adapting the test methods to modificatons made in the user interfaces.

\section{Background and Related Work}

This section provides an introduction to Android application testing, and a survey of existing papers about the challenges it presents.


\emph{Mobile apps} are defined \cite{muccini-1} as mobile software (i.e., applications that run on mobile devices) taking input from the context where they are executed (for instance, contextual sensing and adaptation, and context-triggered actions). They can be distinguished between native apps, if they are designed to run on a specific mobile platform according to its design patterns, or web-based apps, if they are based on web sites engineered to be loaded by a browser application, with partial or no use of the specific functionalities of the mobile device \cite{kirubakaran-1}.

\subsection{Testing Android apps}

Mobile testing can be defined as ``testing native and Web applications on mobile devices using well-defined software test methods and tools to ensure quality in functions, behaviours, performance, and quality of service''\cite{gao-1} .

Testing of mobile apps can be performed on a series of different levels: in addition to the traditional unit testing, integration testing, system testing and regression testing, scopes that are specific to the mobile scenario must be considered. In \cite{kaur-1} compatibility testing (i.e., to ensure that the application works on different handheld models and/or OS versions), performance testing (i.e., to ensure that the mobile devices do not consume too many of the resources available) and security testing are discussed. GUI testing is identified as a very prominent testing need for all mobile applications. For Android applications, GUI testing is focused on testing the \emph{Activities} (i.e. the components in charge of managing the graphical user interfaces) and the transitions between the screens they are composed from.

The first and most immediate option for testing Android applications and their GUIs is the execution of manual test cases. In \cite{vasquez-2}, a study conducted in the field of performance testing, manual testing is identified as the option preferred by developers, along with an examination of reports and feedback from users. The technique, as discussed in \cite{kropp-1}, is however not exhaustive, error prone and not reproducible. 

The approaches for automated GUI testing of Android applications can be classified as follows \cite{linares-1}: fuzzy (or random) testing, model-based testing techniques, capture and replay, white-box scripted testing. Most of them allow, in some cases without having any access to the source code (i.e., only the .apk package of the application is needed), to generate test scripts that can be therefore executed quickly and repeatedly. 

Without any additional information about the AUT (Application Under Test) random and fuzzy testing techniques give random sequences of inputs to activities, in order to trigger potential defects and crashes. Monkey\footnote{https://developer.android.com/studio/test/monkey.html} is the random tester supported by Android. 
Random testers can be applied after a model of the user interface is created (like it is done in \cite{machiry-1}, \cite{moran2017crashscope} and \cite{zhauniarovich-1}) to distribute the input given to the interface in a more intelligent way.

Model-based testing techniques leverage models (typically Finite State Machines or Event-Flow Graphs) of the GUI of the apps under test, that can be created manually or extracted automatically with a process called GUI ripping. Such models are therefore used to generate systematic test cases traversing the GUI. The tools and studies in \cite{amalfitano-5},  \cite{amalfitano-2}, \cite{yang-1} can serve as examples of this approach.

Capture \& Replay testing tools (examples are presented in \cite{gomez-1}, \cite{kaasila-1} and \cite{liu-1}) record the operations performed on the UI to generate repeatable test sequences. Event-sequence generation tools are based on the construction of test cases as streams of events, that then can be inserted in repeatable scripts: \cite{choi-1} and \cite{jensen-1} are examples of this paradigm.

Less coverage (two examples are given in \cite{kropp-1} and \cite{singh-1}) is present in literature about white-box approaches and scripted testing techniques, which require the developer to have access to the code and manually write down testing code with sequences of operations to be performed on the AUT.

Several studies (like \cite{kaur-1} and \cite{muccini-1}) are focused on the peculiarities of Android apps that make testing them properly a complex challenge: limited energy, memory and bandwidth; rapid changes of context and connectivity type; constant interruptions caused by system and communication events; the necessity to adapt the input interface to a wide set of different devices; very short time to market; very high multitasking and interaction with other apps.


The authors in \cite{understanding-culture} find that time constraints, compatibility issues, complexity and lack of documentation of available testing tools are among the most relevant challenges experienced by the interviewed developers, that may therefore be discouraged from testing their applications.

\subsection{Test Fragility}

Test fragility (defined for not GUI-based testing by Garousi et al. \cite{garousi-1}) represents a problem for different kind of software: Leotta et al.~\cite{leotta-1,leotta-2} report a study on web application UI tests. A list of the possible causes of fragilities for mobile applications is reported in~\cite{coppola-1}: identifier and text changes inside the visual hierarchy of activities; deletion or relocation of their elements; usage of physical buttons; layout and graphics change; adaptation to different hardware and device models; activity flow variations; execution time variability.

For our purposes, which is an evaluation of GUI testing of Android apps, we will use the following definition of fragile GUI tests.

\begin{quote}
 A GUI test class is said to be fragile when:
 \begin{itemize} 
  \item it needs modifications when the application evolves;
  \item the need is not due to the modification of the functionalities of the application, but to changes in the interface arrangement and/or definition.
 \end{itemize}
\end{quote}

Modifications performed in test code may be due to different reasons and therefore divided into four categories \cite{yusifo-1}: perfective maintenance, when test code is refactored to enhance its quality (e.g. to increase coverage or to adopt well-known test patterns); adaptive maintenance, to make test code evolve according to the evolutions of the production code; preventive maintenance, to change aspects of the code that may require intervention in future releases; corrective maintenance, to perform bug fixes. According to our definition of GUI testing fragility, we are interested in cases of adaptive maintenance, in which the modifications in the production code are GUI-related.

The manual identification of test fragility occurrences in the history of software projects is time consuming and requires a careful inspection of different version of the test code together with the application production code. 
For those reasons we propose an automatic classification approach: any time a pre-existent method  of a GUI test class is modified we assume the change is due to test fragility. Other test class modifications are not attributed to fragility; for instance, the modifications may involve only import statements and class constructors, or the addition and removal of test methods.
We suppose, in fact, that the addition of a new method should reflect the introduction of new functionalities or new use cases to be tested in the application, and not the modification of existing elements of the already tested activities. On the other hand, if some lines of code inside a single test method had to be changed or added, it is more likely that tests had to be modified due to minor changes in the application and possibly in its user interface (e.g. modifications in the screen hierarchy and in the transitions between activities).



\section{Study Design}

The goals of this work can be described following the Goal-Question-Metric template~\cite{wohlin2012experimentation}: the main objective is to estimate and assess the quantity of fragility-induced changes in test code, in the context of automated scripted GUI testing of Android applications. The goal entails answering the following research questions:

\begin{description}
\item[RQ1]\emph{Diffusion: how many projects use automated testing tools, and how much test code do they produce?}

\item[RQ2]\emph{Evolution: how much test code is modified over different releases?}

\item[RQ3]\emph{Fragility: how fragile are Android UI tests?}
\end{description}

The first step of our research was to estimate the diffusion of Android UI testing. We started from a repository of Android open-source applications -- we selected GitHub for this purpose -- and we performed a code search in order to detect the usage of a set of six testing tools that are frequently cited in literature. 

Then, we studied how applications (and their test classes) were changed throughout their release history, by means of file-by-file comparisons. Finally, with the aid of an automated shell script, we tracked the modifications of individual test classes and methods to compute a set of change indicators. The script cycles over all the releases of each project, for each of them performing the respective \emph{git clone} command to locally download all the files. Then the files are locally investigated to compute size statistics, about Project and Test code. The \emph{git diff} command is leveraged by the script to compute modification statistics between each pair of consequent releases of the project history. Metrics are then computed as explained in detail in section 3.3.

Since we are interested in the evolution of test cases and classes, we excluded from our analysis the projects featuring less than two tagged releases (including master).

Finally, to validate the accuracy of the fragility measures, we selected a random sample of the analyzed projects, and checked the precision of the measures compared to the outcome of a manual inspection of the modified test classes.

\subsection{Metrics definition}
\label{sec:metricsdef}

\begin{table}
 \centering
 \scriptsize
 \caption{Metrics definition}
   \setlength\tabcolsep{2pt}
 \begin{tabular}{lrl}
 \toprule
  \textbf{Group} & \textbf{Name} & \textbf{Explanation} \\
  \hline
  \multirow{5}{*}{\parbox{2cm}{Diffusion and size\\ (RQ1)}} & TD & Tool Diffusion \\
  & NTR & Number of Tagged Releases \\
  & NTC & Number of Test Classes \\
  & TTL & Total Test LOCs \\
  & TLR & Test LOCs Ratio \\
  \hline
  \multirow{5}{*}{\parbox{2cm}{Test evolution\\ (RQ2)}} & MTLR & Modified Test LOCs Ratio \\
  & MRTL & Modified Relative Test LOCs \\
  & TMR & Test Modification Relevance Ratio \\
  & MRR & Modified Releases Ratio \\
  & TSV & Test Suite Volatility  \\
  \hline
  \multirow{7}{*}{\parbox{2cm}{Fragility \\(RQ3)}} & MCR & Modified Test Classes Ratio \\
  & MMR & Modified Test Methods Ratio \\
  & FCR & Fragile Classes Ratio \\
  & RFCR & Relative Fragile Classes Ratio \\
    & FRR & Fragile Releases Ratio \\
  & ADRR & Releases with Added-Deleted Methods Ratio \\
  & TSF & Test Suite Fragility \\
  \bottomrule
 \end{tabular}
 \label{table:metrics_name}
\end{table}

We defined a set of metrics that can be divided into three groups according to the research question they address. Table \ref{table:metrics_name} reports the metrics togehter with the relative descriptions. The metrics are explained in detail in the following subsections.

13 out of the 17 metrics we defined are normalized, to allow comparison across projects of different sizes.
Most of them can be defined on top of lower-level metrics for the quantification of absolute changes in test classes and test cases. For instance, Tang et al.~\cite{tang-1} report eighteen basic metrics for the description of bug-fixing change histories (e.g., number of added or removed files, classes, methods or dependencies). 


\subsubsection{Diffusion and size (RQ1)}

To estimate the diffusion of Android automated UI testing tools and of the size of test suites using them, we defined the following five metrics: 

\begin{description}[leftmargin=0.2cm]

\item[TD] (Tool Diffusion) is defined as the percentage, among the set of Android projects in our context, of those featuring a given testing tool. 

\item[NTR] (Number of Tagged Releases) is the number of tagged releases of an Android project (i.e., the ones that are listed by using the command \emph{git tag} on the GIT repository). 

\item[NTC] (Number of Test Classes) is the number of test classes featured by a release of an Android project, relatively to a specific tool.

\item[TTL] (Total Test LOCs) is the number of lines of code that can be attributed to a specific testing tool 
in a release of an Android project. 

\item[TLR] (Test LOCs Ratio) defined as 
%
$\textit{TLR}_i = \textit{TTL}_i/\textit{Plocs}_i$
%
where $Plocs_i$ is the total amount of Program LOCs for release $i$. This metric, lying in the $[0,1]$ interval, allows us to quantify the relevance of the testing code. 


\end{description}

\subsubsection{Test suite evolution (RQ2)}

The metrics addressing RQ2 aim to describe the evolution of Android projects and the relative test suites; they have been computed for each pair of consecutive tagged releases.

\begin{description}[leftmargin=0.2cm]

\item[MTLR] (Modified Test LOCs Ratio) defined as 
%
$ \textit{MTLR}_i = \textit{Tdiff}_i/\textit{TTL}_{i-1}, $
%
where $\textit{Tdiff}_{i}$ is the amount of added, deleted or modified test LOCs between tagged releases $i-1$ and $i$, and $\textit{TTL}_{i-1}$ is the total amount of test LOCs in release $i-1$. This quantifies the amount of changes performed on existing test LOCs for a specific release of a project. 

\item[MRTL] (Modified Relative Test LOCs) defined as 
%
$ \textit{MRTL}_i = \textit{Tdiff}_i/\textit{Pdiff}_i, $
%
where $\textit{Tdiff}_{i}$ and $\textit{Pdiff}_{i}$ are the amount of modified (or added, or deleted) test and project LOCs, in the transition between release $i-1$ and $i$. It is computed only for releases with test code (i.e., $\textit{TRL}_{i} > 0$). 
This metric lies in the $[0,1]$ range. Values close to 1 imply that a significant portion of the total effort in making the application evolve is needed to keep test methods up to date.

\item[TMR] (Test Modification Relevance Ratio) defined as 
%
$ \textit{TMR}_i = \textit{MRTL}_i/\textit{TLR}_{i-i}. $
%
This ratio can be an indicator of the proportion of effort needed to adapt test classes during the evolution of the application. It is computed only when $\textit{TLR}_{i-1} > 0$. We consider a value greater than 1 as an index of greater effort needed in modifying the test code than the actual relevance of test code inside the application. 

 \item[MRR] (Modified Releases Ratio), computed as the ratio between the number of tagged releases in which at least a test class has been modified, and the total amount of tagged releases. This metric lies in the range $[0,1]$ and bigger values indicate a minor adaptability of the test suite -as a whole- to changes in the AUT.

 \item[TSV] (Test Suite Volatility), is defined for each project as the ratio between the number of test classes that are modified at least once in their lifespan, and the total number of test classes of the project history.

\end{description}

\subsubsection{Fragility of tests (RQ3)}

With an automatic inspection of test code, information about modified methods and classes can be obtained. Based on such data, the metrics answering RQ3 aim to give an approximated characterization of the fragility of test suites.

The number of modified classes with modified methods can be different from the total number of modified classes in three different cases (and their combinations): (i) when the modifications performed to the classes involve non-significant portions of code like comments, imports, declarations; (ii) when the modifications performed to the classes involve only additions of test methods; (iii) when the modifications performed to the classes involve only removal of test methods. Additions and removals of test methods are considered the consequence of a new functionality or a new use case of the application, hence they are not considered as an evidence of fragility of test classes. On the other hand, modifications of test methods may be strictly linked with fragilities.

\begin{description}[leftmargin=0.2cm]
 \item[MCR] (Modified test Classes Ratio) defined as 
 $ \textit{MCR}_i = \textit{MC}_i/\textit{NTC}_{i-1}, $
 where $\textit{MC}_i$ is the number of modified test classes in the transition between release $i-1$ and $i$, and $\textit{NTC}_{i-1}$ the number of test classes in release $i-1$ (the metric is not defined when $\textit{NTC}_{i-1} = 0$). 
 The metric lies in the $[0,1]$ range: the larger the values of $MCR$, the less test classes are stable during the evolution of the app.
 \item[MMR] (Modified test Methods Ratio) defined as 
 $ \textit{MMR}_i = \textit{MM}_i/\textit{TM}_{i-1}, $
 where $\textit{MM}_i$ is the number of modified test methods between releases $i-1$ and $i$, and $\textit{TM}_{i-1}$ is the total number of test methods in release $i-1$ (the metric is not defined when $\textit{TM}_{i-1} = 0$). 
 The metric lies in the $[0,1]$ range: the larger the values of $\textit{MMR}$, the less test methods are stable during the evolution of the app they test.

  \item[FCR] (Fragile Classes Ratio) defined as 
 %
$  \textit{FCR}_i = \textit{MCMM}_i / $ $\textit{NTC}_{i-1}, $
%
 where $\textit{MCMM}_i$ is the number of test classes that are modified, and that feature at least one modified method between releases $i-1$ and $1$. The metric is not defined when $\textit{NTC}_{i-1} = 0$. This metric represents an estimate of the percentage of fragile classes, upon the entire set of test classes featured by a tagged release of the project. The metric is upper-bounded by $\textit{MCR}$, since by its definition $\textit{MCR}_i = \textit{MC}_i / \textit{TC}_i$, and $ \textit{MCMM}_i \le \textit{MC}_i $.

 \item[RFCR] (Relative Fragile Classes Ratio) defined as
 %
$  \textit{RFCR}_i = \textit{MCMM}_i/\textit{MC}_i, $
 where $\textit{MCMM}_i$ and $MC_i$ are defined as above.

 \item[FRR] (Fragile Releases Ratio), computed as the ratio between the number of tagged releases featuring at least a fragile class, and the total amount of tagged releases featuring test classes. This metric lies in the range $[0,1]$ and is upper-bounded by $\textit{MRR}$.
 \item[ADRR] (Releases with Added-Deleted Methods Ratio), computed as the ratio between the number of tagged releases in which at least a test method has been added or removed, and the total amount of tagged releases featuring test classes. This metric lies in the range $[0,1]$, and higher values should imply more frequent changes in application functionalities and defined use cases to be tested.
 
 %
 %

 \item[TSF] (Test Suite Fragility), is defined for each project as the ratio between the number of test classes that feature fragilities at least once in their lifespan, and the total number of test classes of the project history.
 
  \end{description}

To validate the metrics defined for fragile classes and fragile methods (since we may consider as fragile tests that are modified for reasons different from GUI modifications) we adopt the following metric:

\begin{description}[leftmargin=0.2cm]

 \item[P] (Precision), 
  is defined as 
%
$ P = \textit{TP}/(\textit{TP} + \textit{FP}), $
where $\textit{TP}$ is the number of True Positives, in our case the test classes (or methods) that feature changed test code, and whose modifications reflect changes in the GUI of the AUT; $\textit{FP}$ is the number of False Positives, i.e., the test classes (or methods, according to which is being validated) that feature changed test code, but due to different reasons. $\textit{P}$ is defined in the range $[0, 1]$: values closer to 1 are an evidence that the presence of modified lines in test methods is a dependable proxy to identify modifications in test classes due to changes related to the user interface of the application. As our oracle for the computation of Precision, we leverage a manual inspection of a set of selected test classes, before and after they undergo modifications.
\end{description}

\subsection{Selected Testing Tools}
\label{sec:selected-tools}

We have chosen six different popular scripted testing tools for our investigations. We selected open-source testing tools that were already considered in similar explorations of the testing procedure of Android applications. All those testing tools give the possibility to write test scripts manually.

The first two tools we have searched for are part of the official \emph{Android Instrumentation Framework}\footnote{https://developer.android.com/studio/test/index.html}. \emph{Espresso} \cite{knych2014android}
 is an open-source automation framework that allows to test the UI of a single application, leveraging a gray-box approach (i.e., the developer has to know the internal disposition of elements inside the view tree of the app, to write scripts exercising them).  \emph{UI Automator}\cite{linares-1, choudhary2015automated}
adds some functionalities to those provided by Espresso: it allows to check the device status and performance, to perform testing on multiple applications at the same time, and operations on the system UI. Both tools can be used only to test native applications.

\emph{Selendroid}\footnote{https://github.com/selendroid/selendroid} \cite{tan2016research} is a testing framework based on Selenium, that allows to test the UI of native, hybrid and web-based applications; the tool allows to retrieve elements of the application and to inspect the current state of the app's UI without having access to its source code, and to execute the test methods on multiple devices at the same time.

\emph{Robotium}\cite{robotiumbook, grgurina2011development} is an open-source extension of JUnit for testing Android apps, that has been one of the most used testing tools since the beginning of the diffusion of Android programming; it can be used to write black-box test scripts or function tests (if the source code is available) of both native and web-based apps.

\emph{Robolectric}\footnote{http://robolectric.org/} \cite{amalfitano2012toolset, milano2011android, mirzaei2012testing} is a tool that can be used to perform black-box testing directly on the Java Virtual Machine, without the use of a real device or an emulator; it can be considered as an enabler of Test-Driven Development for Android applications, since the instrumentation of Android emulators is significantly slower than the direct execution on the JVM.

\emph{Appium}\cite{singh-1, shah2014software} leverages WebDriver and Selendroid for the creation of black-box test methods that can be run on multiple platforms (e.g., Android and iOS); test methods can be created via an inspector that enables basic functions of recording and playback, via image recognition, or via code. It can be used to test both native and web-based applications. Test scripts can be data-driven.

\subsection{Procedure}

Three main phases can be identified in the study, each relative to one of the three research questions defined. The following paragraphs describe the steps performed in detail.

\subsubsection{Test code Search (RQ1)}

The approach we adopted for the selection of the context (i.e., the set of projects that we used for the subsequent study) is a sequence of different steps, the first one being a search for the word ``Android'' in descriptions, readmes and names of projects. The Repository Search API of Git has been leveraged to this purpose. All the projects extracted this way were cloned locally.

All the projects that have no tagged releases are cut out from the context. This is done because the aim of the experiment is to track the evolutions of the projects, by means of computing differences between tagged releases (as it is explained later). That considered, projects without at least a single tagged release (which allows for a single comparison, made between it and the master release) are not of interest. To know how many releases were featured by each cloned repository, we leveraged the \emph{Git tag} command, which outputs the names of all the tagged releases.

The keyword ``Android'' alone would include libraries, utilities, and applications intended to interface with Android counterparts. 
Since it is mandatory for any Android app to have a Manifest file in its root directory. 
We excluded projects that do not contain any manifest file. 

Once a filtered list of Android projects is obtained, they are searched for the presence of JUnit test classes. The amount of JUnit test classes can serve as a comparison to evaluate the diffusion of other tools and testing techniques. To search for the considered testing tools, a GitHub Code Search, with the names of the tools as keywords, has been performed on the remaining repositories. For each tool its adoption has been estimated by means of the TD metric. Sets of projects featuring different testing tools are not disjoint: it is possible that a repository features more than just one scripted testing tool. Even though some of the chosen tools are based on JUnit, the researches have been conducted independently and in parallel. Obviously, if a tool is based on JUnit, the set of projects featuring JUnit will be a superset of the set of projects featuring that specific tool. The data extraction has been performed between September and December 2016.

We consider any ``.java'' file featuring the name of a testing technique in its code as a test class (for instance, a class featuring the statement ``import static android.support.test.espresso. Espresso.onView;'' is considered as a class featuring Espresso). For each test class the lines of test code are counted, so that \emph{TTL} and \emph{NTC} can be computed for each project, on the master release. The use of the \emph{git tag} command allows to obtain the \emph{NTR} metric.

\subsubsection{Test LOCs analysis (RQ2)}

To answer RQ2, for each pair of consecutive tagged releases of any project, the total amount of modified LOCs is computed. 

Then, the total amount of LOCs added, removed or modified in the test files previously identified is computed. Throughout all our study, we have considered moved or renamed files as different test files.

Those values allow to compute \emph{TLR}, \emph{MTLR}, \emph{MRTL} and \emph{TMR} for each tagged release of the project.

Finally, when the exploration of the project history is complete, global averages are computed: $\overbar{TLR} = Avg_i\{TLR_i\}$, $\overbar{MTLR} = Avg_i\{MTLR_i\}$, $\overbar{MRTL} = Avg_i\{MRTL_i\}$, $\overbar{TMR} = Avg_i\{TMR_i\}$ with $i \in [2, NTR]$, being $NTR$ the number of tagged releases featured by the project. 

Volatile classes (i.e., classes featuring modifications throughout their lifespan) have been identified inside each project, in order to compute the $TSV$ value.

\subsubsection{Test classes history tracking, Fragility (RQ3)}

We have finally tracked the evolution of single test classes and methods, taking into account the tagged releases in which each test class has been added, modified or deleted. 

Then, for each tagged release we have obtained the number of modified classes and methods, i.e. $MCR$ and $MMR$, and the derived metrics $RFCR$ and $FCR$. Also in this case, at the end of the exploration averages have been computed as $\overbar{MCR} = Avg_i\{MCR_i\}$, $\overbar{MMR} = Avg_i\{MMR_i\}$, $\overbar{FCR} = Avg_i\{FCR_i\}$, with $i \in [1, NTR]$.

Since $RFCR$ makes sense only when modifications are actually present, $\overbar{RFCR}$ has been computed as an average of $RFCR$ only for release transitions in which test classes have been modified (i.e., $MCR \ne 0$).

At the end of the exploration of the tagged releases of each project, $\textit{FRR}$ and $\textit{ADRR}$ have been computed to quantify the percentage of them featuring, respectively, fragile and non-fragile modifications.

Based on the recognition of classes affected by fragilities, the overall $\textit{TSF}$ value has been computed for each project.


A manual inspection of a set of modified test classes with modified methods has been conducted, in order to verify the dependability of the metrics defined to identify fragile methods and fragile classes (i.e., $MMR$ and $FCR$). 

30 pairs of consecutive releases of different classes have been selected randomly, and manually inspected before and after they were modified. The modifications performed were characterized under three categories: (i) test code refactoring, syntactical correction and formatting; (ii) adaptation to changes in program code not related to GUI; (iii) adaptation to changes in program code related to GUI.

Only the modifications belonging to the last category are considered as true positives for our analysis; the others are considered as false positives. Based on that subdivision, the precision of the metrics is computed for the percentage of fragile classes, and the percentage of fragile methods.

\section{Results and Discussion}

In the following paragraphs, we report the results we obtained by applying the procedure described in the previous section. The results measured for the metrics defined in section \ref{sec:metricsdef} are detailed, along with the conclusions we can base on them. The full set of intermediate data about classes and releases of each project has been made available online\cite{figshare-dataset}.

We initially gathered a total of 280,447 GitHub repositories featuring the term \emph{Android} in their names, descriptions or readmes. Then, a significant amount of projects were pruned because of their lack of tagged releases (so they had no history to be investigated), or Manifest files. A final set of 18,930 Android projects was obtained (6.75\% of the initial number of projects).

\subsection{Diffusion and size (RQ1)}

%
%
%
%
%

Table  \ref{tablerq1-2} summarizes the metrics gathered to answer RQ1. The columns show: the total number of projects featuring each of the six tools considered; 
the TD metric; the average and median values for $NTR$, $NTC$, $TTL$ and $TLR$, computed on the sets of projects featuring each testing tool. 

As a comparison for the diffusion of other testing tools, we counted the number of projects featuring the JUnit testing framework. We counted 3,669 projects (with tagged releases and manifest files) featuring JUnit, among the total set of Android projects we extracted (the 19.38\%).

\begin{table}[t]
 \centering

 \caption{\emph{NTR}, \emph{NTC}, \emph{TTL}, \emph{TLR} per testing tool: average and median (in parentheses) values for master release.}
 \scriptsize
 \setlength\tabcolsep{4pt}
 \begin{tabular}{@{}lrrr@{~}rr@{~}rr@{~}rr@{~}r@{}}
 \toprule
  Tool & n & TD & \multicolumn{2}{c}{\emph{NTR}} & \multicolumn{2}{c}{\emph{NTC}} & \multicolumn{2}{c}{\emph{TTL}} & \multicolumn{2}{c}{\emph{TLR}}\\
  \hline
  Espresso & 423 & 2.23\% & 15 &(6) & 5 &(2) & 588 &(190) & 8.8\% &(4.1\%) \\
  UIAutomator & 134 & 0.71\% & 60 &(25) & 12 &(3) & 3,155 &(1,134) & 8.6\% &(0.6\%)\\
  Selendroid & 6 & 0.03\% & 46 &(17) & 76 &(1) & 8,627 &(126) & 19.4\% &(0.2\%) \\
  Robotium & 150 & 0.79\% & 44 &(7) & 5 &(1) & 873 &(227) & 8.7\% &(3.3\%) \\
  Robolectric & 842 & 4.44\% & 22 &(6) & 11 &(3) & 1,448 &(399) & 16.4\% &(11.4\%) \\
  Appium & 18 & 0.09\% & 27 &(15) & 38 &(4) & 4,469 &(1096) & 37.3\% &(6.0\%)\\
  \bottomrule
 \end{tabular}
 \label{tablerq1-2}

\end{table}

Considering the overestimation due to possible overlaps (since the sets for the individual tools are not necessarily disjoint) about 8.5\% of the set of projects feature tests belonging to one of the six selected tools. None of the testing frameworks reached by itself a significant level of diffusion. The absolute number of projects featuring Selendroid and Appium test classes is practically irrelevant. A higher number (the 4.44\% of the total) of projects featuring Robolectric has been found, but the tool has been available for a longer time with respect to other ones (especially Espresso and UI Automator) and is often used solely for Unit Testing.

Although the total number of Android projects extracted can take into account some projects that are not likely to feature test classes (e.g. experiments, duplicates, exercises, prototypes, projects that are abandoned at very early stages) the statistics extracted about the metric $TD$ give evidence of the lack of an extensive usage of scripted automated UI testing on Android. However, it must be taken into account that the study we performed is limited to the testing tools we considered, i.e. it is possible that different scripted testing tools are used by some other projects of the context.

The average and median number of test classes can be quite small (e.g., just 5 and 2, respectively, in the case of Espresso) due to the typical coding patterns for Android applications, in which -usually- one GUI testing class is written specifically for each Activity featured by the application. Most applications -this is particularly evident in the case of small and even experimental open-source projects- do not feature many screens to be shown to their users, and therefore they do not feature many activities to be tested.

Average $TTL$ and $TLR$ values are very large for both Selendroid and Appium; however, the result is heavily influenced by the small size of the sets of projects featuring these tools (respectively 6 and 18 projects) and by the presence of the full Selendroid framework for Android (selendroid/selendroid, with 47,436 LOCs) and of a very large set of Appium API demos (appium/android-apidemos, with 48,868 LOCs). 

The fact that the set of projects featuring Espresso has the lowest average $TTL$ can be explained with the following reasons: (i) using a white-box testing technique allows to exercise the functionalities of the application with little coding effort; (ii) the framework is quite accessible even to non-experienced developers, and its usage is encouraged by Android, leading it to be used also in very small projects, in tryouts, and even for experimental and partial coverage of applications use cases. On the other hand, the mean $TTL$ for projects featuring UI Automator is very high, and also significantly higher with respect to the sets featuring Robotium, Robolectric and Espresso. This is mainly due to the cross-application features of UIAutomator, that make it recommended for the testing of whole firmwares and application bases, which are typically very big projects.\\[0.5ex]

\noindent\fbox{\begin{minipage}{\dimexpr\columnwidth-2\fboxsep-2\fboxrule\relax}
The considered GUI testing tools reach a diffusion that is always lower than 4.5\%. Projects that have their GUI tested feature on average 9 test classes, with a total of 1,361 LOCs (13.2\% of the whole project code).
\end{minipage}}

\subsection{Test suite evolution (RQ2)}

Table \ref{table:evolution} shows the statistics collected about the average evolution of test code, for the six selected testing tools. For every set, $\overbar{TLR}$, $\overbar{MTLR}$, $\overbar{MRTL}$, $\overbar{TMR}$, $MMR$ and ${TSV}$ have been averaged on all the projects. The values in last row are obtained as averages of the six values above, weighted by the respective sizes of the six sets.

The values reported for average Test LOCs Ratio ($\overbar{TLR}$) show that -when present- GUI testing can be an important portion of the project during its lifecycle, if compared to the number of LOCs of program code. The average values range from about 7.3\% (for the set of Espresso projects) to 31.9\% (for the set of Appium projects). For the largest set of projects considered (the ones featuring Robolectric) the mean $\overbar{TLR}$ is 13.4\%. 

Average Modified Test LOCs Ratio ($\overbar{MTLR}$) measures show that typically around 2.8\% of test code is modified between consecutive releases. Very small values were obtained for the projects featuring UIAutomator. In general, this should be a consequence of bigger test suites, in terms of absolute LOCs, with respect to the ones written with other testing frameworks. Hence, the influence of a similar amount of absolute modified LOCs would result in a lower $\overbar{MTLR}$ value. The highest value was found for the set of projects featuring Selendroid: this can be explained by the very high percentage of total LOCs belonging to testing code for these repositories. However, the set of projects featuring Appium did not exhibit the same trend, having a lower $\overbar{MTLR}$: this should mean that, even though the important ratio of testing code above project code, few modifications (in both production and test code) were made between subsequent releases.

\begin{table}
\centering
 \scriptsize

\caption{Measures of the evolution of test code (averages on the sets of repositories)}

\label{table:evolution}
 \setlength\tabcolsep{4pt}
 \begin{tabular}{lrrrrrr}
 \toprule
  Tool & \emph{$\overbar{TLR}$} & \emph{$\overbar{MTLR}$} & \emph{$\overbar{MRTL}$} & \emph{$\overbar{TMR}$}& $MRR$ & TSV\\
  \hline
  Espresso & 7.3\% & 2.6\% & 4.7\% & 0.68& 22.2\% & 28.6\% \\
  UI Automator & 9.6\% & 1.4\% & 3.5\% & 1.17& 16.5\% & 35.9\%\\
  Selendroid &  19.4\% & 4.3\% & 11.5\% & 0.15 & 39.6\% & 33.7\%\\
  Robotium & 7.8\% & 3.8\% & 5.3\% & 0.56 & 22.1\% & 36.3\%  \\
  Robolectric & 13.4\% & 2.9\% & 9.5\% & 0.79 & 28.2\% &  30.4\% \\
  Appium  & 31.9\% & 1.8\% & 16.6\% & 0.27 & 27.3\% & 36.2\% \\
  \hline
  Average & 11.1\% & 2.8\% & 7.4\% & 0.76 & 25.2\% & 30.6\% \\
  \bottomrule
 \end{tabular}
\end{table}

The measures about Modified Relative Test LOCs ($\overbar{MRTL}$) show that, on average, when UI testing tools are used, the 7.4\% of the modified LOCs belong to test classes. With this metric, however, we are still unable to discriminate what is the reason behind the modifications to be performed on test classes. The higher $\overbar{MRTL}$ values for the sets of projects featuring Appium and Selendroid can be justified by the small size of the two sets, and by the nature of the projects examined. For instance, the Selendroid framework, on GitHub as selendroid/selendroid, is subject to heavy modifications.

The mean values of Test Modification Relevance Ratio ($\overbar{TMR}$) stayed in the range between 0.56 and 1.17 for big-sized sets of projects, with lower values for sets featuring Selendroid and Appium. In general, those values imply that the effort to spend in modifying test code is not linear with the relevance of test code inside the application: in our case, on average, the ratio between the intervention on test code and the intervention on program code is about 3/4 of the ratio between test and program code. The higher $\overbar{TMR}$ value for UIAutomator is due to some projects (e.g. Lanchon/android-platform-tools-base) in which $TLR$ is rather small, and where in some releases all modified LOCs belong to test classes (thus leading to $MRTL$ values very close to 1).

The Modified Releases Ratio (${MRR}$) metric gives an indication about how often the developers had to modify any of their test classes when they published new releases of their projects. On average, 25.2\% of releases needed modifications in the test suite (with a maximum of 39.6\% for the set of projects featuring Selendroid). Since releases may be frequent and numerous for GitHub projects, this result explains that the need for updating test classes is a common issue for Android developers that are leveraging scripted testing. The average 30.6\% value for the Test Suite Volatility (TSV) metric, which characterizes the phenomenon from the point of view of whole test suites, highlights that on the lifespan of a project, about one third of test classes require at least one modification. \\[0.2ex]

\noindent\fbox{\begin{minipage}{\dimexpr\columnwidth-2\fboxsep-2\fboxrule\relax}
On average, near 3\% of testing code is modified between consecutive tagged releases. 7.4\% of the overall LOCs modified between consecutive tagged releases belong to testing code. On average, one fourth of tagged releases require modifications in the test suite, and one third of the test suites needs modifications during the project history.
\end{minipage}}

\subsection{Fragility of tests (RQ3)}

Table \ref{table:fragility} shows the fragility estimations that we have computed for each project, and then averaged over the six sets: $\overbar{MCR}$, $\overbar{MMR}$, $\overbar{FCR}$, $\overbar{RFCR}$. Based on them, we computed three additional derived metrics: $FRR$, $ADRR$ and ${TSF}$. The values in last row are obtained as averages of the six values pertaining to the individual sets of projects, weighted by the respective sizes of the six sets.



The first column about the Modified Classes Ratio ($\overbar{MCR}$) metric shows that, on average, 14.8\% of test classes are modified between consecutive tagged releases in our set of Android projects. The only value significantly different from the average is the one obtained for the set of projects featuring UIAutomator, but it can be justified with the bigger amount of test classes that they feature on average (see table \ref{tablerq1-2}).

The 3.6\% average value found for the Modified Methods Ratio ($\overbar{MMR}$) metric highlights that the percentage of modified methods is -as expected- smaller than the percentage of modified classes: this is obviously due to the fact that multiple test methods are contained in single test classes. 

Not all modified test classes could be defined as fragile classes. The Relative Fragile Classes Ratio ($\overbar{RFCR}$) metric gives a statistic about the possibility of a modified class to contain modified methods. The results collected show that more than half of the classes having modified lines featured modifications inside the code of test methods as well, hence they could be defined as fragile according to the heuristic definition given in section 2.4. The Fragile Classes Ratio ($\overbar{FCR}$) metric gives the ratio between the classes that we define fragile upon all the classes contained by each project. On average, 8.2\% of the classes were fragile in the transition between consecutive releases of the same project.

The Fragile Releases Ratio ($FRR$) metric gives an indication of how many releases of the considered project contained test classes that we identify as fragile. The value is upper-bounded by $MRR$, which is the frequence of releases featuring any kind of modification. The average value for $\overbar{FRR} = 17.7\%$ means that about one every five releases records fragility-induced changes in test methods. The Releases with Added-Deleted Methods Ratio ($ADRR$) metric quantifies the probability that there is the need -- between two subsequent releases -- to add or delete test methods inside existing test classes. In general (with the only exception of the set of projects featuring UIAutomator) $ADRR$ is higher than $FRR$. 
This result is in accordance with the findings by Pinto et al.\cite{pinto-1}, who observed -- in the context of traditional desktop applications -- that the sum of test deletions and additions is higher, on average, than the number of test modifications.
However, we can observe that the two values are generally close to each other: during the evolution the need for fragility induced test changes ($FRR$) occurs roughly as often as the definition of new test methods ($ADRR$).

Upper-bounded by ${TSV}$ (the overall volatility for test suites), the average value for Test Suite Fragility ($TSF$) provides information about the amount of test classes, in each project, that need modifications because of fragilities. The average value of 20.2\% tells us that one fifth of the classes in test suites face at least a fragility during its entire lifespan.\\[0.5ex]

\begin{table}
\centering
\scriptsize

\caption{Measures for RQ3 (averages on the sets of repositories)}
\label{table:fragility}
 \setlength\tabcolsep{3.5pt}
 \begin{tabular}{lrrrrrrr}
 \toprule
  {\bf Tool} & $\overbar{MCR}$ & $\overbar{MMR}$ & $\overbar{FCR}$ & $\overbar{RFCR}$ & $FRR$ & $ADRR$ & $TSF$\\
  \midrule
  Espresso & 15.2\% & 3.5\%& 8.3\% & 59.7\% & 14.4\% & 17.7\% & 18.8\%\\
  UI Automator & 9.0\% & 1.8\%& 4.6\%  & 54.4\% & 10.2\% & 8.2\% &16.6\%\\
  Selendroid & 16.5\% & 2.7\%& 4.9\% & 42.2\% &28.2\%& 23.2\% &11.9\%\\
  Robotium & 16.4\% & 3.5\%& 9.3\% & 53.1\% & 15.2\%& 21.2\% & 22.8\%\\
  Robolectric & 15.1\% & 3.8\% & 8.5\%& 60.7\% & 20.6\% & 25.8\% & 19.4\%\\
  Appium & 15.2\% & 4.6\%  & 7.7\%& 48.2\% & 17.1\% & 23.5\%  & 19.6\% \\
  \midrule
  \emph{Average} & 14.8\% & 3.6\%  & 8.2\%& 59.1\% & 17.7\%& 21.9\% & 20.2\%\\
  \bottomrule
 \end{tabular}
\end{table}


\noindent\fbox{\begin{minipage}{\dimexpr\columnwidth-2\fboxsep-2\fboxrule\relax}
On each new release 14.8\% of test classes and 3.6\% of test methods are modified. Fragility-induced changes concern 8.2\% of the classes. One every five releases feature fragile test classes, and 20.2\% of the classes inside test suites are affected by fragilities at least once in their lifespan.

Overall the changes induced by fragility requires an effort comparable to the definition of tests for new features: both in terms of frequency (17.7\% of releases with fragility-induced changes vs. 21.9\% of releases with new or removed test methods) and number of classes interested (among the modified test classes 59.1\% are affected by fragilities).
\end{minipage}}\\[0.5ex]

It must also be considered that the averages are significantly lowered by projects in which test classes have been added -- at the beginning or at some point in their history -- but never modified:
in practice tests fell into oblivion. 
For instance, among the projects of 423 projects featuring Espresso, we detected modifications in test classes in 181 projects (43\%), and modifications in test method in 144 projects (34\%).


\subsection{Fragility Metrics Validation}


Table \ref{table:validation} shows the results of the validation procedure for RQ3, described in section 3.3.3. 
We found that about 69\% of the modifications of methods are true positives if we consider them as proxies of modifications performed to the GUI. Hence, we can consider that modifications in the GUI of the AUT are involved in the majority of the modifications to test methods and classes. 

Considering fragility at class level -- i.e. classes containing at least a fragile method --, we found 21 such classes, and hence 21 true positives among 30 samples (70\%).
Such classification performance is comparable to that achieved with widely adopted approaches for fix-commit identification~\cite{AntoniolFix2008}.\\[0.1ex]


\noindent\fbox{\begin{minipage}{\dimexpr\columnwidth-2\fboxsep-2\fboxrule\relax}
In 70\% of the samples analyzed, a modification in a test method (or class) corresponds to an actual GUI test fragility being addressed.
\end{minipage}}

\begin{table}[hbt]
 \centering
 \small
 \caption{Precision for Fragile Methods and Classes}
 
 \label{table:validation}
 
 \begin{tabular}{@{}lrrrr@{}}
  \toprule
  Metric & Measured & TP & FP & P\\
  \hline
  Fragile Methods & 65 & 45 & 20 & 69\%\\
  Fragile Classes & 30 & 21 & 9 & 70\%\\
  \bottomrule
  
 \end{tabular}

\end{table}

\section{Threats to Validity}

\emph{Threats to internal validity}. 
%
 The test class identification process is based on the search of the name of the tool as keyword: any file containing one of such keyword is considered as a test file without further inspection; this procedure may miss some test classes, or consider a file as a test file mistakenly.
The number of tagged releases is used as a criterion to identify a project as worth to be investigated; it is not assured that this check is the most dependable one for pruning negligible projects.
The release level has been selected as the granularity of our inspections. We have considered that the commit level would have been an excessively fine granularity, taking in consideration very small and/or temporary modifications, and non-relevant dynamics. However, average metrics computed on the release level can be slightly different from the ones computed on the commit level, and it may be the case that changes between releases cancel each other out.
The scripts and tools we used assume that no syntactic errors are present inside the test classes on which they operate, and that the names of those files are properly spelled (e.g., without the presence of special characters or blank spaces); the correctness of the metric extraction technique is not assured in different circumstances.

\emph{Threats to external validity}. 
%
Our findings are based only on the GitHub open-source project repository. Even though it is a very large repository, it is not assured that such findings can be generalized to closed-source Android applications, neither to ones taken from different repositories. The applications we extracted are not necessarily released to final users. Nevertheless, we selected a subset of projects that were released on the Play Store, and the average metrics computed on them were not significantly different from the ones computed on the whole sample.
We have collected measures for six scripted GUI automated testing tools. It is not certain that such selection is representative of other categories of testing tools or different tools of the same category, which may exhibit different trends of fragilities throughout the history of the projects featuring them.

\emph{Threats to construct validity}. We link the GUI test fragility to any change in the interface that requires an adaptation of the test. 
The proxy we used - a change in any test method - is not perfectly linked  to a change in the GUI. The magnitude of this threat has been evaluated with a Precision measure equal to 70\%. This might reduce our fragility estimate but not change its order of magnitude.


\section{Conclusion and Future Work}

In this paper we aimed at taking a snapshot of the usage of automated GUI testing frameworks in the Android ecosystem. We analyzed the use of some of the most important tools -- Espresso, UI Automator, Selendroid, Robotium, Robolectric, and Appium -- in the projects hosted on the GitHub portal.

The level of adoption of any GUI testing framework is about the 8\% of the Android projects having at least one tagged release. This value can be compared to the 20\% diffusion for the JUnit framework in the same context. Overall, automated GUI testing is not widely adopted. This result is slightly lower of the one about the F-Droid repository by Kochar et al.~\cite{understanding-culture}, who found that only 14\% of apps contained test classes, and only 9\% of apps had executable test classes. On average, when present, the GUI testing code represent about 11\% of the whole project code. 

Concerning the evolution of test code, in each release, on average, about 7.5\% of the changed lines are in the GUI test code and about 3\% of test code is modified. 

The fragility of the tests can be estimated with two metrics based on the raw count of classes and methods modified. Overall we can estimate fragility of the analyzed test classes around 8\% (meaning that there is such probability that a test class may include a modified test method). On average, one out of five classes in each test suite needs modifications in its code because of fragilities. The association between modified test methods and fragility has been proved dependable in 70\% of the samples examined. These results show that developers need rather frequently to adapt their GUI scripted testing suites, and suggest that state of the art tools should profit of additional features reducing the amount of effort needed by users to keep their script up to date and running. The results can also be used as a benchmark for practitioners and developers.

Based on these evaluations and insights about the fragility issue, we plan as future work to define a taxonomy of the causes of fragilities, produce a set of actionable guidelines to help developers avoiding them, and finally develop automated tools capable of adapting the test classes and methods to modificatons made in the user interfaces. An extension of the study to other databases of open-source projects, to take into account different testing frameworks or to other software platforms (like iOS) is also possible.\\[1ex]

\textbf{Acknowledgment.} This work was supported by a fellowship from TIM.

\bibliographystyle{ACM-Reference-Format}
\bibliography{sigproc} 

\end{document}